\documentclass[10pt,letterpaper,twocolumn]{article} 

\usepackage{ol2}
\usepackage[draft]{hyperref}
\usepackage{amsmath}

\begin{document}

\twocolumn[ 

\title{Invisible non-Hermitian potentials in discrete-time photonic quantum walks}


\author{Stefano Longhi}
\address{Dipartimento di Fisica, Politecnico di Milano and Istituto di Fotonica e Nanotecnologie del Consiglio Nazionale delle Ricerche, Piazza L. da Vinci 32, I-20133 Milano, Italy (stefano.longhi@polimi.it)}
\address{IFISC (UIB-CSIC), Instituto de Fisica Interdisciplinar y Sistemas Complejos, E-07122 Palma de Mallorca, Spain}

\begin{abstract}
Discrete-time photonic quantum walks on a synthetic lattice,  where both spatial and temporal evolution of light is discretized,  have provided recently a fascinating platform for the observation of a wealth of non-Hermitian physical phenomena and for the control of light scattering in complex media. A rather open question is whether 
invisible potentials, analogous to the ones known for continuous optical media, do exist in such discretized systems.
 Here it is shown that, under certain conditions, slowly-drifting Kramers-Kronig potentials behave as invisible potentials in discrete-time  photonic quantum walks. 
\end{abstract}

 ] 

{\em Introduction.} The scattering of waves through inhomogeneous or disordered systems is ubiquitous in different areas of classical and quantum physics. In optics, light scattering arises in any inhomogeneous medium where the refractive index rapidly changes on a spatial scale of the order of the optical wavelength. 
However, it is known since long time that some special inhomogeneous distributions of the refractive index do not reflect light \cite{r1}. Recently, there has been a 
surge of interest in controlling the scattering of waves through inhomogeneous or disordered media based on the special engineering of both the real and imaginary parts of the refractive index \cite{r2,r3,r4,r5,r6,r7,r7b,r8,r9,r10,r11,r12,r13,r14,r15,r16,r17,r18,r19,r19b,r20,r21,r22,r23,r24,r25,r26,r27}.
 Refractive index engineering enables to realize 
new kinds of reflectionless and even invisible potentials, such as those based on parity-time (PT) symmetry and transformation optics \cite{r2,r3,r4,r5,r6,r7}, reverse engineering \cite{r7b,r8,r9} and the spatial Kramers-Kronig relations \cite{r10,r11,r12,r13,r14,r15,r16,r17,r18,r19,r19b,r20}. Complex potential engineering offers as well a systematic method to construct a complex medium where a desired waveform can freely propagate, free of scattering, even in highly-disordered media  \cite{r21,r22,r23,r24,r25,r26,r27}.
However, the experimental feasibility to engineer at sub-wavelength scale both real and imaginary parts of the refractive index in continuous media is challenging, and there are few experiments demonstrating such new classes of synthetic materials either at microwaves \cite{r19} or for acoustic waves \cite{r23}.\\  Discrete-time photonic quantum walks on a lattice \cite{r28} provide a different and important class of optical systems, where the evolution of light is discretized both in space and time. As compared to continuous optical media, they offer the rather unique possibility of a simple implementation of complex (non-Hermitian) potentials with rather arbitrary profiles, providing a fascinating platform to experimentally access a wealth of novel non-Hermitian phenomena, such as PT symmetry breaking \cite{r29,r30}, non-Hermitian topological physics \cite{r31,r32,r33,r34,r34b}, non-Hermitian Anderson localization \cite{r35} and non-Hermitian phase transitions \cite{r36}. Recently, the observation of constant-intensity waves and induced transparency has been reported using discrete-time photonic quantum walks with complex tailored potentials \cite{r37}, thus overcoming the limitations of refractive index engineering required in continuous media. However, an open question is whether in such discrete-time systems there exist potentials that are invisible for any incident wave, i.e. not only for a given target  wave shape. This question is specially demanding since the discrete nature of the dynamics can deeply modify the scattering properties of non-Hermitian reflectionless potentials known in continuous media \cite{r18}.\\
In this Letter we consider wave scattering in a discrete-time photonic quantum walk setting from non-Hermitian complex potentials that drift on the lattice at a constant speed, and show that the class of Kramers-Kronig potentials are invisible in the slowly-drifting regime.\\
 \begin{figure}
   \centering
    \includegraphics[width=0.45\textwidth]{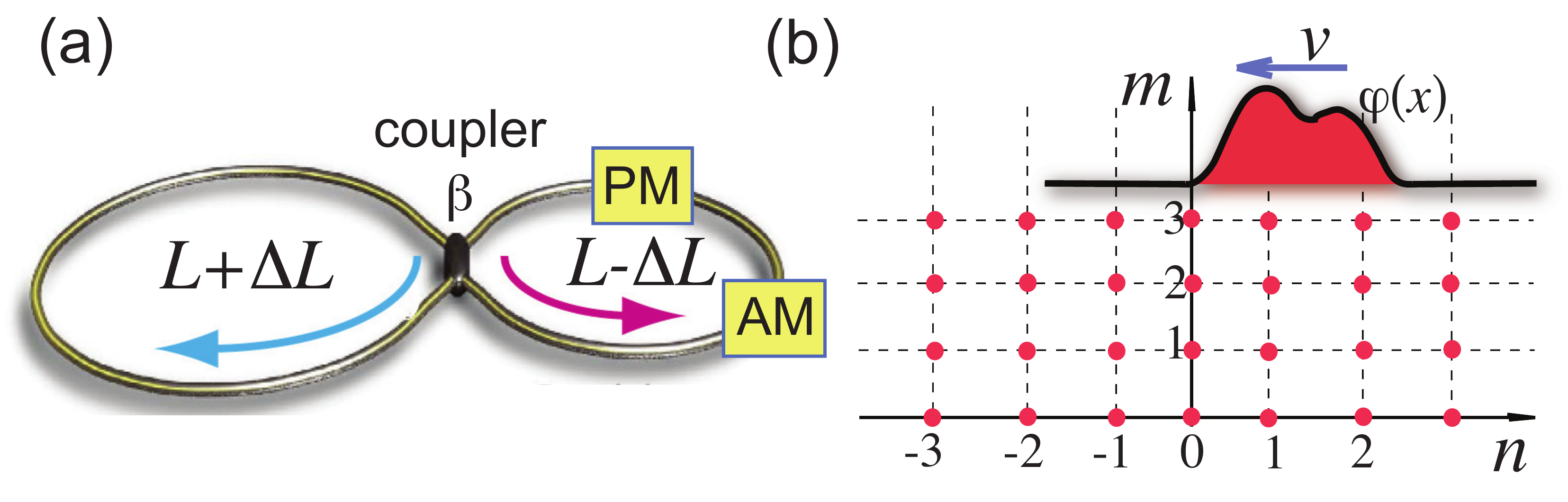}
    \caption{Discrete-time photonic quantum walk on a synthetic lattice with a non-Hermitian scattering potential. (a) Schematic of two coupled fiber loops with slightly length mismatch $L \pm \Delta L$. A fiber coupler with coupling angle $\beta$ mixes the light waves between the two fiber loops. An amplitude (AM) and a phase (PM) modulator are placed in the short loop, providing an effective complex potential $V_{n,m}$ for the traveling pulses in the loop.  (b) Schematic of the synthetic mesh lattice. The physical time $t$ is mapped at the discretized times  $t_n^m=n \Delta T+mT$, where $T=L/c$ is the mean travel time and $\Delta T= \Delta L /c \ll T$ is the travel time mismatch between the two fiber loops. The index $n$ corresponds to the site index in a synthetic one-dimensional spatial lattice, while the integer $m$ corresponds to a discrete time along which  the system evolves. The scattering potential is assumed of the form $V_{n,m}= \varphi(n+mv)$, i.e. drifting along the lattice in the backward direction at a speed $v$.}
\end{figure}
\\
{\it Model.}  We consider a discrete-time photonic quantum walk realized using optical pulses in a synthetic mesh lattice \cite{r28,r29,r30,r33,r35,r36,r37}. The system  consists of two fiber loops of slightly different lengths  $L \pm \Delta L$ (short and long paths) that are connected by a fiber coupler with a coupling angle $\beta$ [Fig.1(a)]. In the short loop, we place an amplitude (AM) and a phase (FM) modulator, which provide a desired control of both phase and amplitude of the traveling pulses. Light dynamics in the fiber loops mimic a one-dimensional discrete-time quantum walk of a single quantum particle \cite{r28}
with time steps $m$ and spatial positions $n$ [Fig.1(b)]. Neglecting pulse broadening and cross-talk effects, the coupled equations for the amplitudes
$u_n^{(m)}$ and $v_n^{(m)}$ of the optical pulses in the short and long fiber loops read \cite{r29,r30,r35,r37,referee}
 \begin{eqnarray}
 u^{(m+1)}_n & = & \left[   \cos \beta u^{(m)}_{n+1}+i \sin \beta v^{(m)}_{n+1}  \right]  \exp (-iV_{n,m+1})  \\
 v^{(m+1)}_n & = & \left[   \cos \beta v^{(m)}_{n-1}+i \sin \beta u^{(m)}_{n-1}  \right] 
 \end{eqnarray}
where $V_{n,m}$ is the discrete complex potential that is generated by the combination of AM and PM modulators. {The ratio $ L / \Delta L= T / \Delta T $ determines the number of sites in the lattice, whereas the travel time mismatch $\Delta T=\Delta L /c$ gives the upper limit to the duration $\Delta \tau$ of pulses that can be injected in the loops \cite{referee}. Typical values of mean travel time $T=L/c$ and time mismatch $\Delta T$ are $T \sim 27 \; \mu$s and $\Delta T \sim 50$ ns \cite{r36}, so that the lattice can accommodates more than $200$ sites. For a typical pulse duration  $ \Delta \tau\sim 5$ ns, pulse broadening due to fiber dispersion is negligible over several thousands of time steps $m$, so that the mesh lattice equations (2) describe with excellent accuracy the pulse dynamics in the two rings (see e.g. \cite{r30,r33,r36}).} 

 In the absence of the potential, i.e. for $V_{n,m}=0$, the system shows discrete translational invariance both in space and time, and the eigenfunctions are of the Bloch-Floquet form, i.e. $(u_n^{(m)},v_n^{(m)})^T=(\bar{U}_{\pm}, \bar{V}_{\pm})^T \exp[iqn-i E_{\pm}(q) m]$, where $q$ is the (spatial) Bloch wave number,
\begin{equation}
E_{\pm}(q) = \pm {\rm acos} ( \cos \beta \cos q)
\end{equation}
\begin{figure}
   \centering
    \includegraphics[width=0.45\textwidth]{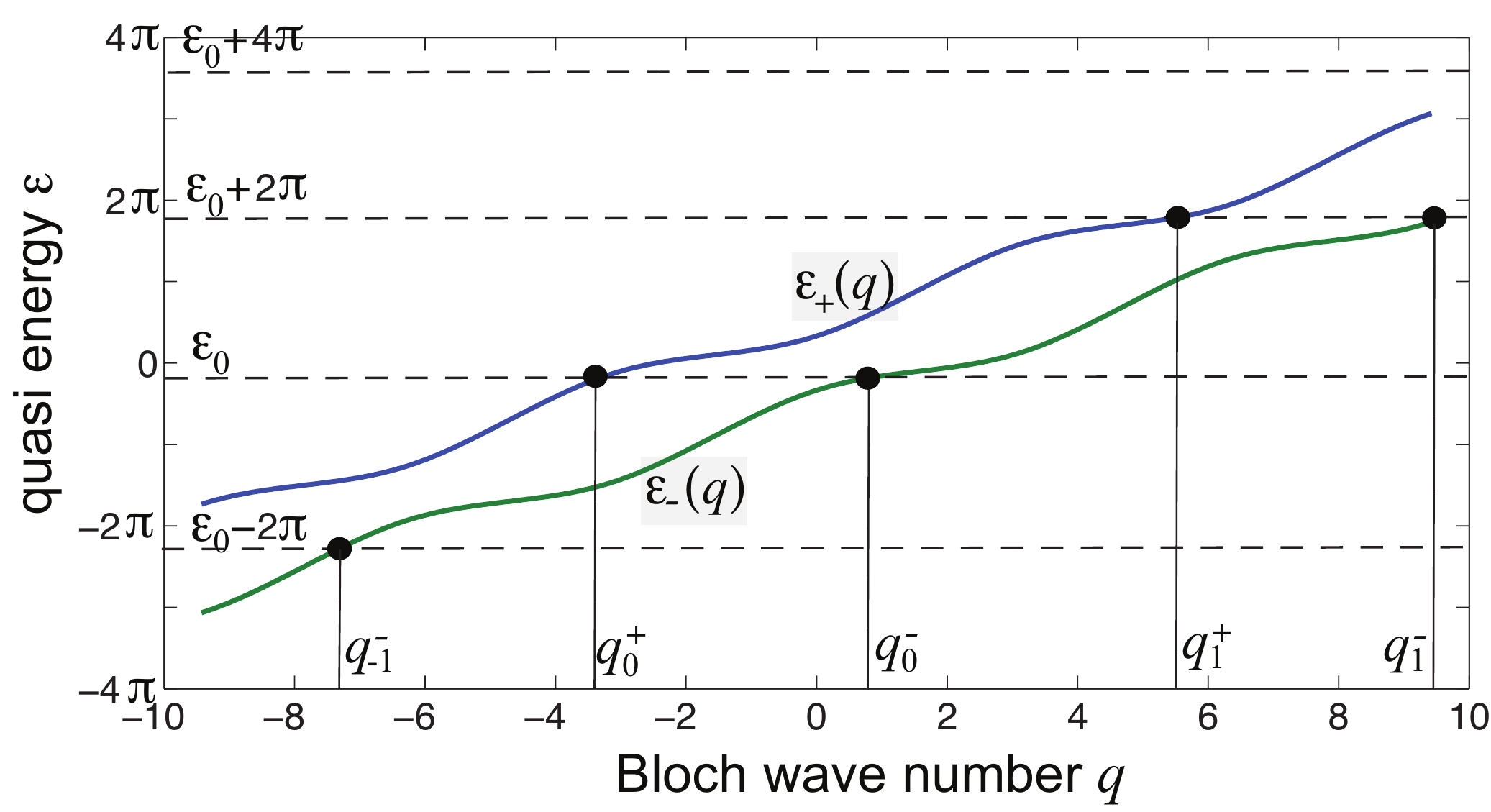}
    \caption{Behavior of the quasi energies $\epsilon_{\pm}(q)$ versus Bloch wave number $q$ (solid curves) for the discrete-time photonic quantum walk in the moving reference frame $(x,t)$ [Eq.(7)]. Parameter values are $\beta= \pi/3$ and $v=0.8> \cos \beta$. For an incoming wave with Bloch wave number $q_0^+$ of energy $\epsilon_0=\epsilon_+(q_0^+)$ belonging to the upper quasi energy band, the scattered wave contains all Bloch wave numbers $q_{\alpha}^{\pm}$ of various scattering channels, defined by the equations $\epsilon_{\pm}(q^{\pm}_{\alpha})= \epsilon_0+ 2 \pi \alpha$, with $\alpha=0, \pm1, \pm2, ...$.}
\end{figure}
are the quasi-energies of the two bands of the binary lattice, and $\bar{U}_{\pm}=i \sin \beta \exp(iq)$, $\bar{V}_{\pm}=\exp(-i E_{\pm})- \cos \beta \exp(iq)$ are the amplitudes of Bloch waves. Note that a wave packet with carrier Bloch wave number $q$ in either one of the two bands travels in the lattice at the speeds (group velocites) $v_{g \pm}=(d E_{\pm} /dq)$, which take the largest absolute value $v_{g}^{(max)}= \cos \beta$ at $q= \pm \pi/2$. Therefore, any excitation in the discrete lattice cannot propagate faster than $v_{g}^{(max)}$. \\
\\
\begin{figure}[h]
  \centering
    \includegraphics[width=0.46\textwidth]{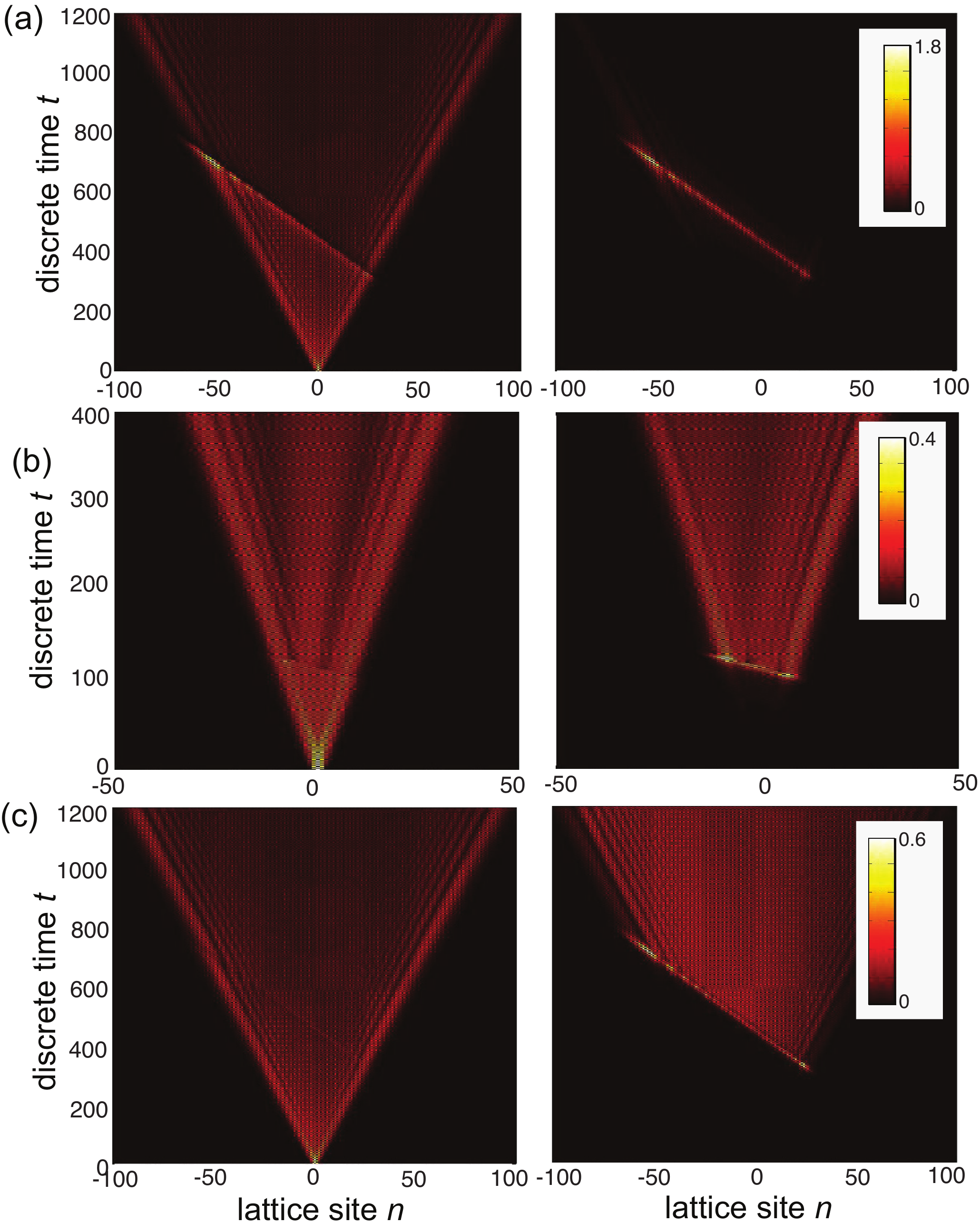}
    \caption{(a) Scattering dynamics from the Kramers-Kronig potential $\varphi(x)=A/(x-x_0)^2$ ($A=-i$, $x_0=90+i$) for a coupling angle $\beta= 0.95 \times \pi/2$ and drift velocity $v=0.2$. The initial excitation of the lattice is $u_n^{(0)}=v_n^{(0)}= \delta_{n,0}$. The left (right) panel shows the discrete dynamics of 
    $P_n^{(m)}$ ($Q_n^{(m)}$) on a pseudo color map, where $P_n^{(m)}=|u_n^{(m)}|^2+|v_n^{(m)}|^2$ and  $Q_n^{(m)}=|u_n^{(m)}-\bar{u}_n^{(m)}|^2+|v_n^{(m)}-\bar{v}_n^{(m)}|^2$. Here  $\bar{u}_n^{(m)}$ and $\bar{v}_n^{(m)}$ are the pulse amplitudes that would propagate in the lattice in the absence of the scattering potential. The vanishing of $Q_n^{(m)}$ after the scattering event clearly indicates that the potential is invisible. (b) Same as (a), but for a drift velocity $v=0.8$. In this case the fast-moving potential is not anymore invisible. (c) Same as (a), but for a Hermitian scattering potential $\varphi(x)={\rm Re} \left( A/(x-x_0)^2 \right)$.}
\end{figure}
{\it Wave scattering analysis.} Let us consider the discrete-time photonic quantum walk in the presence of a space-time complex potential $V_{n,m}$, localized in space and vanishing fast enough as $n \rightarrow \pm \infty$. We aim at establishing whether one can find a class of potentials that are fully invisible for any arbitrary wave packet propagating in the system, i.e. such that after the scattering event the wave packet evolves exactly as if the potential were not present. We note that such a kind of transparency has been recently demonstrated in Ref.\cite{r37}, however in that case the invisibility holds only for a target incident waveform. Additionally, we mention that the families of static reflectionless potentials known in continuous media, such as the class of Kramers-Kronig potentials \cite{r13}, may loose their reflectionless property  owing to space discretization \cite{r18}.\\
To search for a class of invisible potentials, let us assume, for the sake of definiteness, that the wave packet comes from $n=-\infty$ and propagates in the forward direction of the lattice. Since in the clean system there is an upper bound $v_g^{(max)}$ to the propagation speed of excitation, it readily follows that any potential  of the form $V_{n,m}= \varphi(n+mv)$, i.e. drifting on the lattice at a speed $v$ in the backward direction [Fig.1(b)], is reflectionless provided that $v>v_g^{(max)}= \cos \beta$ \cite{r38}: in fact, any scattered wave cannot appear as a reflected wave in the reference frame where the potential is at rest. Here $\varphi(x)$ is a continuous function of the variable $x$, that defines the shape of the scattering potential. The absence of reflected waves, however, does not ensure that the potential is invisible. To study the scattering dynamics, it is worth considering the moving reference frame 
\begin{equation}
x=n+vm , \;\; t = m 
\end{equation}
where the potential is at rest. Note that in the moving reference frame the variable $t$ remans discrete, while $x$ should be considered as a continuous variable. After letting $f(x,t)=u_{x-vt}^{(t)}$ and $g(x,t)=v_{x-vt}^{(t)}$, the discrete dynamics in the $(x,t)$ frame reads
\begin{eqnarray}
f(x,t+1) & = &  [ \cos \beta f(x-v+1,t)+ i \sin \beta g(x-v+1,t)] \nonumber \\
& \times & \exp[-i \varphi(x)] 
\end{eqnarray}
\begin{equation}
g(x,t+1)  =  i \sin \beta f(x-v-1,t)+ \cos \beta g (x-v-1,t).
\end{equation}
Note that, in the absence of the scattering potential [$\varphi(x)=0$], the Bloch-Floquet eigenstates of the system are given by $(f(x,t),g(x,t))^T=(\bar{F}_{\pm}, \bar{G}_{\pm})^T \exp[iqx-i \epsilon_{\pm}(q)t]$, where the quasi energies $\epsilon_{\pm}(q)$ in the moving reference frame reads [compare with Eq.(3)]
\begin{equation}
\epsilon_{\pm}(q)=E_{\pm}(q)+qv=qv \pm {\rm acos} ( \cos \beta \cos q).
\end{equation}
and where $(\bar{F}_{\pm}(q),\bar{G}_{\pm}(q))^T=( i \sin \beta \exp[iq(1-v)], \exp(-i \epsilon_{\pm}) - \cos \beta \exp[iq(1-v)])^T$ are the amplitudes of Bloch waves. A typical diagram of the quasi energies $\epsilon_{\pm}(q)$ is shown in Fig.2. Note that, when the condition $v> \cos \beta$ is satisfied, the curves $\epsilon_{\pm}(q)$ are monotonously increasing functions of $q$, indicating that in the moving reference frame there cannot be reflected (backward-propagating) waves. In the presence of the scattering potential, we need to consider the scattering solutions to Eqs.(5) and (6) with the asymptotic form corresponding to plane waves, with a given Bloch wave number $q=q_0^+$ of the incident wave. In order to ensure that the scattering waves far from the potential region are of Jost type (i.e. plane waves), we assume that $\varphi(x)$ vanishes as $x \rightarrow \pm \infty$ faster than $ \sim 1/x$ \cite{r11}. For the sake of definiteness, let us assume that the incoming Bloch wave belongs to the upper band of the lattice, so that its quasi energy is $\epsilon_0= \epsilon_+(q_0^+)$ and 
$(f(x,t),g(x,t))^T \simeq (\bar{F}_+(q_0^+),\bar{G}_+(q_0^+))^T \exp(iq_0^+x-i \epsilon_0 t)$ as $x \rightarrow - \infty$. The same analysis holds if one assumes an incoming wave belonging to the lower band. Since in the scattering process the quasi energy is conserved apart from integer multiples than $2 \pi$, the asymptotic solution of the scattered wave as $ x \rightarrow \infty$ must be of the form
\begin{equation}
\left(
\begin{array}{c}
f(x,t) \\
g(x,t)
\end{array}
\right) \simeq \sum_{\alpha, \pm} t_{\alpha}^{\pm} 
\left(
\begin{array}{c}
\bar{F}_{\pm}(q_{\alpha}^{\pm}) \\
\bar{G}_{\pm} (q_{\alpha}^{\pm}) 
\end{array}
\right)
\exp (iq_{\alpha}^{\pm}x -i \epsilon_0 t)
\end{equation}
with some amplitudes $t_{\pm}^{\alpha}$, where $\alpha=0, \pm 1, \pm 2,..$ is the order of scattering channel and $q_{\alpha}^{\pm}$ are the roots of the equation $\epsilon_{\pm}(q_{\alpha}^{\pm})=\epsilon_0+ 2 \pi \alpha$ (see Fig.2). Basically, the amplitudes $t_{\alpha}^{\pm}$ are the transmission coefficients of various scattering channels in the two bands, labelled by the index $\alpha$ and arising from the discrete nature of time evolution. Clearly, the scattering potential is invisible provided that all amplitudes $t_{\alpha}^{\pm}$ vanish, apart from $t_0^+$ which should be $t_0^+=1$. The analytic form of $t_{\alpha}^{\pm}$ can be derived in the weak potential limit $|\varphi(x)| \ll 1$ using a first-order (Born) approximation, and turns out to be proportional to $\hat{\varphi}(q_{\alpha}^{\pm}-q_0^+)$, where $\hat{\varphi}(q)= \int dx \varphi(x) \exp(-iq x)$ is the Fourier spectrum of the potential (Sec.I of the Supplemental document). Therefore, it is not possible rather generally to have an invisible potential, owing to the infinite number of scattering channels. However, in the limit of a slowly-drifting potential $v \rightarrow 0$, all the wave numbers $q_{\alpha}^{\pm}$ of scattered waves diverge like $ \sim 1/v$, apart from $q_0^+$ which does not depend on $v$. Note that 
the slowly-drifting regime $v \rightarrow 0$ necessarily implies $\beta \rightarrow \pi /2^- $, with $\cos \beta <v$. Since the Fourier spectrum $\hat{\varphi}(q)$ of the scattering potential vanishes at high spatial frequencies,  in the $v \rightarrow 0$ regime the amplitudes $t_\alpha^{\pm}$ of scattered waves are vanishing, with the exception of $t_0^+$. In the Born approximation and for a slowly-driving potential, invisibility is thus attained provided that $\hat{\varphi}(q=0)=0$. As shown in Sec.II of the Supplemental document, using the method of complex spatial displacement \cite{r14} it can be shown that invisibility is found for any slowly-drifting potential $\varphi(x)$ of the Kramers-Kronig type, i.e. for which its Fourier spectrum $\hat{\varphi}(q)$ vanishes for either $q \geq 0$ or $q \leq 0$, beyond the Born approximation. For example, any potential of the form 
\begin{equation}
\varphi(x)= \sum_l A_l (x-x_l)^{-h_l}, 
\end{equation}
with $A_l$ and $x_l$ arbitrary complex numbers, with the only constraint ${\rm Im}(x_l)>0$ [or ${\rm Im}(x_l)<0$], and $h_l$ integer numbers larger than one, is invisible in the slowly-drifting regime.\\
\\
\begin{figure}
   \centering
    \includegraphics[width=0.48\textwidth]{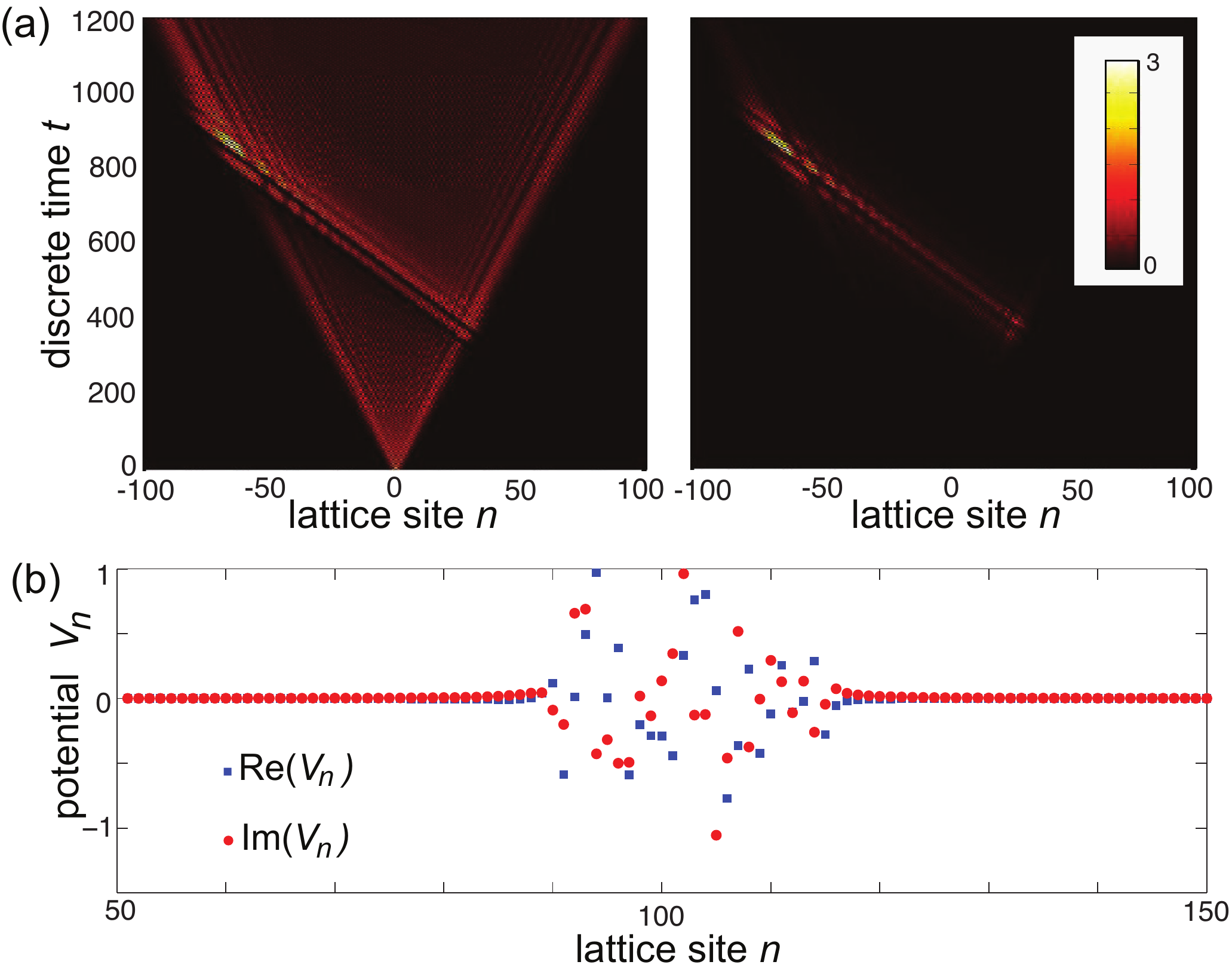}
    \caption{(a) Scattering from a slowly-drifting Kramers-Kronig potential with an irregular profile (drift velocity $v=0.2$, coupling angle $\beta= 0.95 \times \pi/2$). The left (right) panel shows the discrete dynamics of 
    $P_n^{(m)}$ ($Q_n^{(m)}$) on a pseudo color map. (b) The behavior of the potential $V_{n,m}$ (real and imaginary parts) at initial time $m=0$. The potential is obtained from the sum defined by Eq.(9) with 25 terms  with poles $x_l=90+l+i$  and random complex amplitudes $A_l$ [the modulus $|A_l|$ is uniformly distributed in the interval $(-0.5,0.5)$, whereas the phase of $A_l$ is uniformly distributed in the range $(0, 2 \pi)$].}
\end{figure}
{\it Numerical results.} We checked the predictions of the theoretical analysis by direct numerical simulations of the discrete-time equations (1) and (2). The equations have been integrated assuming as a typical initial condition a symmetric single pulse excitation of the lattice at the spatial site $n=0$, i.e. $u_n^{(0)}=v_n^{(0)}=\delta_{n,0}$, far apart from the scattering region, however we checked that the invisibility can be observed for rather arbitrary initial excitation of the lattice. Figures 3(a) and (b) show, as an example, the scattering dynamics from the potential $\varphi(x)=A/(x-x_0)^2$, displaying a single pole of second order, in a a photonic quantum walk with coupling angle $\beta= 0.95 \times \pi /2$ and for two values of the drift velocity $v$. In Fig.3(a), the scattering potential drifts slowly on the lattice and it appears invisible, whereas in Fig.3(b) the potential drifts fast and ceases to be invisible, according to the theoretical analysis. Clearly, the invisibility of the potential is lost even in the slowly-drifting regime when $\varphi(x)$ is not  of the Kramers-Kronig type. This is shown, as an example, in Fig.3(c), where the scattering potential $\varphi(x)$ is real (Hermitian scattering) and given by the real part solely of the potential $A/(x-x_0)^2$. Finally, we note that, as mentioned above, the shape of the Kramers-Kronig potential can be made arbitrarily irregular by considering the multi-pole potential (9) with the sum extended over a suitable large number of terms: in spite of the irregular shape of the potential, it is invisible for any incident wave in the slowly-drifting regime. An example of an irregular multi-pole invisible potential is shown in Fig.4.\\
\\
 {\it Conclusion.} We predicted the existence of invisible potentials of Kramers-Kronig type in discrete-time photonic quantum walks, extending to time-discrete systems the fascinating property of such a class of non-Hermitian potentials, unraveling some limitations arising from the discrete nature of the temporal dynamics. Our results could be readily extended to discrete-time multi-band lattice systems, providing a deeper understanding of non-Hermitian wave scattering and invisibility in discrete-time  systems and opening up new possibilities for non-Hermitian wave control.\\


\begin{thebibliography}{99}


\bibitem{r1}
I. Kay and H. E. Moses, J. Appl.  Phys. {\bf 27}, 1503 (1956).
\bibitem{r2}
Z. Lin, H. Ramezani, T. Eichelkraut, T. Kottos, H. Cao, D. N. Christodoulides, Phys. Rev. Lett. {\bf 106}, 213901
(2011).
\bibitem{r3}
S. Longhi, J. Phys. A {\bf 44}, 485302 (2011).
\bibitem{r4}
L. Feng, Y.-L. Xu, W.S. Fegadolli, M.-H. Lu, J.E.B. Oliveira, V.R. Almeida, Y.-F. Chen, and A. Scherer, Nature Mat. {\bf 12}, 108 (2013).
\bibitem{r5}
A. Mostafazadeh, Phys. Rev. A {\bf 87}, 012103 (2013).
 \bibitem{r6}
  X. Zhu, L. Feng, P. Zhang, X. Yin, X. Zhang, Opt. Lett. {\bf 38}, 2821 (2013).
 \bibitem{r7}
 S.A.R. Horsley, C.G. King, and T.G. Philbin, J. Opt. {\bf 18}, 044016 (2016).
 \bibitem{r7b}
 S. Longhi,
Phys. Rev. A {\bf 82}, 032111 (2010).
 
\bibitem{r8}
F. Loran and A. Mostafazadeh,
Phys. Rev. A {\bf 100}, 053846 (2019).
\bibitem{r9}
Z. Hayran, R. Herrero, M. Botey, H. Kurt, and K. Staliunas, Phys. Rev. A {\bf 98}, 013822 (2018).



\bibitem{r10} 
S.A.R. Horsley, M. Artoni, and  G.C. La Rocca, Nature Photon.{\bf 9}, 436 (2015).
\bibitem{r11} 
S. Longhi,  EPL {\bf 112}, 64001 (2015).
\bibitem{r12} 
S. Longhi, Opt. Lett. {\bf 41}, 3727 (2016).
\bibitem{r13} 
 S.A.R. Horsley, M. Artoni, and G.C. La Rocca, 
Phys. Rev. A {\bf 94}, 063810 (2016).
\bibitem{r14} 
 S.A.R. Horsley and S. Longhi,  Am. J. Phys. {\bf 85}, 439 (2017).
 \bibitem{r15} 
C.G. King, S.A.R. Horsley, and T.G. Philbin, Phys. Rev. Lett. {\bf 118}, 163201 (2017).
\bibitem{r16} 
S.A.R. Horsley and S. Longhi,
Phys. Rev. A {\bf 96}, 023841 (2017).
\bibitem{r17} 
F. Loran and A. Mostafazadeh, Opt. Lett. {\bf 42}, 5250 (2017).
\bibitem{r18} 
S. Longhi, Phys. Rev. A {\bf 96}, 042106 (2017).
 \bibitem{r19}
 D. Ye, C. Cao, T. Zhou, J. Huangfu, G. Zheng, and L. Ran, Nature Commun. {\bf 8}, 51 (2017).
 \bibitem{r19b}
 W.W. Ahmed, R. Herrero, M. Botey, Y. Wu, and K. Staliunas, Phys. Rev. Applied {\bf 14}, 044010 (2020).
\bibitem{r20} 
Y. Zhang, J.H. Wu, M. Artoni, and G.C. La Rocca, 
Opt. Express {\bf 29}, 5890 (2021).
\bibitem{r21}
K.G. Makris, A. Brandst\"otter, P. Ambichl, Z. H. Musslimani, and S. Rotter, Light Sci. Appl.
{\bf 6}, e17035 (2017).
\bibitem{r22}
S. Yu, X. Piao, and N. Park, Phys. Rev. Lett. {\bf 120}, 193902 (2018).
\bibitem{r23}
E. Rivet, A. Brandst\"otter, K.G. Makris, H. Lissek, S. Rotter, and R. Fleury, Nature Phys. {\bf 14}, 942 (2018).
\bibitem{r24}
A. Brandst\"otter, K. G. Makris, and S. Rotter, Phys. Rev. B. {\bf 99}, 115402 (2019).
\bibitem{r25}
A. F. Tzortzakakis, K. G. Makris, S. Rotter, and E.N. Economou, Phys. Rev. A {\bf 102}, 033504 (2020).
\bibitem{r26}
K. G. Makris, I. Kresic, A. Brandst\"otter, and S. Rotter, Optica {\bf 7}, 619 (2020).
\bibitem{r27}
I. Komis, S. Sardelis, Z.H. Musslimani, and K.G. Makris, Phys. Rev. E. {\bf 102}, 032203 (2020).


\bibitem{r28}
 A. Schreiber, K. N. Cassemiro, V. Potocek, A. Gabris, P. J. Mosley, E. Andersson, I. Jex, and
C. Silberhorn,  Phys. Rev. Lett. {\bf 104}, 050502 (2010).
 \bibitem{r29}
 A. Regensburger, C. Bersch, M. A. Miri, G. Onishchukov, D.N. Christodoulides, and U. Peschel,
Nature {\bf 488}, 167 (2012).
 \bibitem{r30}
 M. Wimmer, M. A. Miri, D. Christodoulides, and U. Peschel, Sci. Rep. {\bf 5}, 17760 (2015).
 \bibitem{r31}
X. Zhan, L. Xiao, Z. Bian, K. Wang, X. Qiu, B.C. Sanders, W. Yi, and P. Xue, 
Phys. Rev. Lett. {\bf 119}, 130501 (2017).
\bibitem{r32}
L. Xiao, T. Deng, K. Wang, G. Zhu, Z. Wang, W. Yi, and P. Xue, Nature Phys. {\bf 16}, 761 (2020).
\bibitem{r33}
S. Weidemann, M. Kremer, T. Helbig, T. Hofmann, A. Stegmaier, M. Greiter, R. Thomale, and
A. Szameit, Science {\bf 368}, 311 (2020).
\bibitem{r34}
K. Wang, T. Li, L. Xiao, Y. Han, W. Yi, and P. Xue, Phys. Rev. Lett. {\bf 127}, 270602 (2021).
\bibitem{r34b}
S. Longhi, Opt. Lett. {\bf 47}, 2951(2022).
\bibitem{r35}
S. Weidemann, M. Kremer, S. Longhi, and A. Szameit, Nature Photon. {\bf 15}, 576 (2021).
\bibitem{r36}
S. Weidemann, M. Kremer, S. Longhi, and A. Szameit, Nature {\bf 601}, 354 (2022).
\bibitem{r37}
A. Steinfurth, I. Krexic, S. Weidemann, M. Kremer,
K.G. Makris, M. Heinrich, S. Rotter, and A. Szameit,
 Sci. Adv. {\bf 8 }, eabl7412 (2022).
 \bibitem{referee}
 S. Longhi, Phys. Rev. B {\bf 105}, 245143 (2022).
 \bibitem{r38}
 S. Longhi, Opt. Lett. {\bf 42}, 3229 (2017).



\end{thebibliography}
\end{document}